\documentstyle[12pt]{article}    
\begin{document}
\begin{center}
  
{\bf Brane-Antibrane Systems Interaction under Tachyon Condensation}

\vspace{1cm}

                      Wung-Hong Huang\\
                       Department of Physics\\
                       National Cheng Kung University\\
                       Tainan,70101,Taiwan\\

\end{center}
\vspace{3cm}

      The interaction between a parallel brane-antibrane and brane-antibrane is investigated by regarding the brane-antibrane pair as a kink or anti-kink type tachyon condensed state.   As the kink-type tachyon condensed state is known as a non-BPS brane we expand the Lagrangian of tachyon effective field theory  to the quadratic order in the off-diagonal fluctuation and then use the zeta-function regularization and Schwinger perturbative formula to evaluate the interaction within a kink-kink or a kink-antikink.   The results show that while the kink and kink has repulsive force the kink and anti-kink has attractive force and may annihilate by each others.   We therefore evaluate the free energy at finite temperature and determine the critical temperature above which the stable state of kink-antikink system may be found.

\vspace{4cm}

\begin{flushleft}
   
E-mail:  whhwung@mail.ncku.edu.tw\\

\end{flushleft}

\newpage
\section{Introduction}

    Sen's conjectures on tachyon condensation  [1] have drawn attention to various non-BPS D-brane configurations in string theory [2-6].  According to the conjectures the potential height of the tachyon potential exactly cancels the tension of the original unstable D-brane and, at the stable true vacuum, the original D-brane disappears and closed string theory is realized. Also, the kink-type tachyon condensations correspond to lower dimensional D-branes and organize ``Descent relations'' between stable and unstable D-branes [2].  These conjectures have been examined from various viewpoints in string/M theory, such as boundary string field theories (BSFT) [7].  While the simplified field theory model of tachyon dynamics [3,4] is a powerful tool in describing various aspects of these dynamics in a simpler context.

   The generalization to the brane-antibrane system has also been investigated [5,6].   It is seen that, for a Dp-\={Dp} pair system the kink-like configuration of the tachyon leads to a non-BPS D(p-1) brane [5].  In a recent paper [6], Hashimoto and Sakai used a tachyon effective field theory of a non-BPS brane to construct a  classical solution representing a parallel brane-antibrane. From the solution the energy for the brane and antibrane with distance $x$ is proportional to $~ x^{n} e^{-x^2/4}$.  The exponential factor in here denotes the appearance of an excitation of  a string connecting the two branes.    

   In this paper we use the tachyon effective field theory of a non-BPS brane  [3,4,8] to investigate the interaction between a parallel brane-antibrane and brane-antibrane.    We regard one brane-antibrane pair as a kink and another pair as a kink or an anti-kink configuration.   As the kink or anti-kink in here is corresponding to the tachyon condensed state it could be described by the non-BPS brane.   We therefore expand the corresponding Lagrangian to the quadratic order in the off-diagonal fluctuation to evaluate the interaction within a parallel kink-kink or a parallel kink-antikink.  Using the zeta-function regularization and Schwinger perturbative formula  [9-11] we show that the force between the kink and kink is repulsive.   However, the force between the kink and anti-kink is attractive and thus may annihilate by each others.  We also evaluate the free energy at finite temperature and determine the critical temperature above which the stable state of kink-antikink system may be found.

    Note that the free energy for the brane-antibrane system at finite temperature had been evaluated in [12] in which the D-\={D} system is simply regarded as a free-gas thermal system with classical tachyon field mass +  tachyon classical field potential.   However, our treatments are to consider the interaction between a parallel brane-antibrane and brane-antibrane  which is different from them.

\section{Tachyon Effective Field Theory} 

   Lagrangian of the two derivative truncation of boundary string field theory is given by [3,4,8]

$$L_{D_p}= -\tau _p \left(\frac{1}{2}\partial_\mu T\partial_\mu T + 1\right) e^{-T^2/4}. \eqno{(2.1)}$$ 
where $\tau_p$ is the tension of the unstable $D_p$ brane.  The unstable vacuum at $T=0$ correspond to the original $D_p$ brane, and its energy density is exactly equal to the $D_p$ brane tension.  On other hand, the stable vacuum at $T=\pm\infty$ is thought of as the "closed string vacuum" with vanishing energy.    It is seen that there is another stable one soliton solution with $T=\sqrt{2} x$ which represents a kink solution with center at $x=0$. This kink solution has tension $\tau_{p-1}=2\sqrt{2\pi}\tau_p$ which is  reasonably close to the actual value of $\tau_{p-1}=\sqrt{2}\pi\tau_p$. It was also shown in [3] that the fluctuation modes about this soliton have integer mass squared level spacing.  Thus, this model nicely conforms with the Sen conjectures [1].

\subsection{Lagrangian of  Brane-Antibrane System}

It is known that the kink-type tachyon condensation of a brane-antibrane pair becomes a unstable D-branes [2].   In order to describe kink-kink or kink-antikink system the tachyon field needs to be generalized to a $2$ by $2$ matrix [4,8].   In this paper we use the following Lagrangian of the tachyon effective field   

   $$L=  -\tau_p Tr \left[~ \frac{1}{2} \partial_ \mu \bar{T} e^{-\bar{T}T/8}\partial_\mu T e^{-\bar{T}T/8} + e^{-\bar{T}T/4}~\right].   \eqno{(2.2)}$$
\\
This Lagrangian has been derived from the string theory by Kutasov, Marino and Moore [4].   Minahan [4] had also used this Lagrangian to investigate the property of stretched strings in tachyon condensation models.  Note that in the kink-antikink system the tachyon field become complex-valued and the tachyon potential in the above equation shall take the form $Tr (e^{-T \bar{T}/4})$, as expected.   The tachyon field can be expressed as 

$$T=\left(\begin{array}{cc}\sqrt{2}(x-x_0)&T_+\\T_-&-\sqrt{2}(x+x_0)\end{array}\right)  \eqno{(2.3a)} $$
\\
in which $T_+$=$\bar T_-$ [4,8].    The diagonal parts represent a kink (i.e., a non-BPS brane after the tachyon condensation of a brane-antibrane pair) with the center located at $x_0$ and an antikink with center located at $-x_0$.   The off-diagonal complex tachyon fields are $T_+$ and $T_-$ which represent the lowest energy excitation of the strings stretched between a non-BPS brane (i.e., a kink) and non-BPS antibrane (i.e., an anti-kink).   In a same way the tachyon field corresponding a kink with the center located at $x_0$ and a kink with center located at $-x_0$ is 

$$T=\left(\begin{array}{cc}\sqrt{2}(x-x_0)&T_+\\T_-&\sqrt{2}(x+x_0)\end{array}\right)  \eqno{(2.3b)} $$
\\

   To calculate the Lagrangian we first use the Pauli matrix $\sigma_i$ to express the tachyon field (2.3a) as

$$T= - \sqrt {2} x_0 1 + \sqrt {2} x \sigma_z + {1\over2}(T_++T_-)\sigma_x + {i\over2}(T_+-T_-)\sigma_y.  \eqno{(2.4)}$$
Then, using the formula

$$e^{x \sigma_x + y \sigma_y + z \sigma_z}=cosh(\sqrt{x^2+y^2+z^2})  + {x\sigma_x+y\sigma_y+z\sigma_z\over \sqrt{x^2+y^2+z^2}} sinh(\sqrt{x^2+y^2+z^2}),   \eqno{(2.5)}$$
the Lagrangian expanding to the second order in the off-diagonal tachyon field becomes

$$L_{(K-\bar K)} \approx   - ~[~ 4 K(x,x_0) + K(x,x_0) ~ \partial_\mu T_+ \partial_\mu T_- + U(x,x_0) ~ T_+T_-~].    \eqno{(2.6a)}$$
in which and hereafter we let $\tau_p=1$.    We also define    

$$K(x, x_0) \equiv e^{- (x^2+x_0^2)/2} cosh(xx_0). \hspace{3.5cm} \eqno{(2.7a)}$$
$$ U(x, x_0) \equiv e^{- (x^2+x_0^2)/2} \left({x_0\over x}sinh(xx_0)-cosh(xx_0)\right).  \eqno{(2.7b)}$$
After the integration the first term in the  (2.6) can be expressed as    

$$\int_{-\infty}^{\infty} dx [ e^{- (x^2+x_0^2)/2} ~ cosh(xx_0)] = {1\over 2}\int_{-\infty}^{\infty} dx [ e^{- (x-x_0)^2)/2}+e^{- (x+x_0)^2)/2}] = \sqrt {2\pi}. \eqno{(2.8)}$$
Thus, the first term clearly represent the energy density of the independent kink and antikink, which is irrelevant to our investigation and is neglected hereafter. 

    To find the interaction between kink and antikink we shall integrate out the off-diagonal tachyon field in (2.6).   However, in the above truncated Lagrangian the coefficients before the tachyon kinetic term and mass term, i.e. $K(x, x_0)$ and $U(x, x_0)$, are coordinate dependent and we have to adopt some approximations to proceed.

   Our prescription is to redefined a new  real field $\Phi_i$ by

$$T_\pm = \sqrt{K(x,x_0)}~ (\Phi_1 \pm i  \Phi_2).  \eqno{(2.9)}$$
Then we have the following relation

 $$\int dx ~ K(x,x_0) ~\partial_\mu T_+ \partial_\mu T_-  = - \int dx \left[\Phi_1 ~\partial^2 \Phi_1 + \Phi_2 ~\partial^2 \Phi_2 + M_k(x,x_0) \left(\Phi_1^2+ \Phi_2^2 \right)\right], \eqno{(2.10a)} $$
 $$\int dx ~ U(x,x_0) T_+ T_-  = \int dx~ M_u(x,x_0) \left(\Phi_1^2+ \Phi_2^2 \right), \hspace{6cm} \eqno{(2.10b)} $$
in which 
  $$M_k(x,x_0) \equiv - {1\over4} K(x,x_0)^{-2} \partial_\mu K(x,x_0)\partial_\mu K(x,x_0) + {1\over 2} K(x,x_0)^{-1}  \partial_\mu \partial_\mu K(x,x_0) $$
$$\hspace{0.5cm}= {1\over4} (2-x^2) - {1\over2} (1-x^2) x_0^2 + {1\over 12} (3 -2 x^2) x_0^4 + O(x_0^6),  \hspace{1cm} \eqno{(2.11a)}$$

$$M_u(x,x_0) \equiv {U(x,x_0)\over K(x,x_0)} = -1 + x_0^2 - {1\over 3} x^2 x_0^4 + O(x_0^6). \hspace{3cm}  \eqno{(2.11b)}$$
\\
This means that our approximation is to consider the kink antikink at short distance, i.e. $x_0<1$ (note that we have set $\tau_p=1$ ).    The associated Hamiltonian operator with respect to the fields $\Phi_i$ can be expanded as

$$H=H_0 + V_1  + O(x_0^4), \eqno{(2.12)}$$
$$H_0 \equiv \partial_\mu \partial_\mu +( {3\over2} - {x^2\over4}), \eqno{(2.13a)}$$
$$V_1\equiv -{1\over2}(1+x^2) x_0^2 , \hspace{0.5cm}\eqno{(2.13b)}$$
\\
It is fortunate that the operator $H_0$ has eigenfunction of parabolic cylinder function  $D_n(x)$ [13], i.e.

$${d^2\over dx^2} D_n(x)+ (n + {1\over2} - {x^2\over4})D_n(x) = 0, \eqno{(2.14)}$$
$$D_n(x) = 2^{-n\over 2} e^{-x^2\over 4} (2\pi)^{-{1\over4}} (\sqrt {n!})^{-1}H_n({x\over\sqrt{2}}),\eqno{(2.15)}$$
\\
in which  $H_n$ is the Hermite function.  Note that the above definition renders the parabolic cylinder function  $D_n(x)$ been normalized.

   For the tachyon field (2.3b) the Lagrangian expanding to the second order in the off-diagonal tachyon field becomes

$$L_{(K-K)} \approx   - ~[~ 4 K(x,x_0) + K(x,x_0) ~ \partial_\mu T_+ \partial_\mu T_- ~].    \eqno{(2.16)}$$
\\
and the associated Hamiltonian operator become 
$$H=H_0 + V_1  + O(x_0^4), \eqno{(2.17)}$$
$$H_0 \equiv \partial_\mu \partial_\mu +( {1\over2} - {x^2\over4}), \eqno{(2.18a)}$$
$$V_1\equiv -{1\over2}(1-x^2) x_0^2 , \hspace{0.5cm}\eqno{(2.18b)}$$

   The existence of the exact solution of  operator $H_0$ allows us, as in the conventional quantum field [11], to use the zeta-function regularization method  to evaluate the renormalized effective action with a help of the Schwinger perturbative formula [10].   Using the effective action we can then obtain the interaction and free energy in the kink-antikink systems.   The method is briefly described in below.

\subsection{Effective Action and Schwinger Perturbation Method }

   First, for a field $\Phi$ with action $S[\Phi]$ the effective action $W$ defined by 

$$e^{iW} =\int D\Phi e^{iS[\Phi]}, \eqno(2.19) $$
can be evaluated by the $\zeta$-function regularization [9-11]

$$W =-i{1\over2} ln[Det(H)] = -i{1\over2}[\zeta'(0) + \zeta(0) ln(\mu^2)], \eqno(2.20) $$
We will take parameter $\mu =1$ in this paper.   The $\zeta$ function can be evaluated from the relation

$$\zeta_H(\nu) = [\Gamma(\nu)]^{-1} \int dxdt  \int_0^{\infty}  ids (is)^{\nu -1}<x,t\mid e^{-isH}\mid x,t>,    \eqno(2.21)$$
where the operator H is defined in (2.12) ot (2.16).   Next, as the quantities $V_1$ is small we can use the following expansion [10]
 
 $$Tr e^{-isH} = Tr [ e^{-isH_0} - is e^{-isH_0} V_1  ],                 \eqno(2.22)$$
to expand the $\zeta$ function by

   $$ \zeta_H(\nu)= \zeta(\nu)_0 +\zeta(\nu)_{V_1} + \cdot\cdot\cdot, \eqno(2.23)$$
where $\zeta_ i(\nu)$ are defined in below.

	$$\zeta_0(\nu)= [\Gamma(\nu)]^{-1} \int dx dt  \int_0^{\infty}  ids (is)^{\nu -1}<x,t\mid e^{-isH_0}\mid x,t>,  \eqno(2.24a)$$

 	$$\zeta_{V_1}(\nu)= [\Gamma(\nu)]^{-1} \int dx dt \int_0^{\infty}  ids (is)^{\nu -1}(-is) <x,t\mid e^{-isH_0} V_1\mid x,t>, \eqno(2.24b)$$
\\
In the next we will use the above relations to evaluated the $\zeta$ function and then obtain the energy and free energy in the kink-antikink systems.

\section{Calculations of Zeta-Function and Effective Action} 

   The $\zeta_0(\nu)$ does not depend on $x_0$ and is irrelevant to our discussions.   We can neglect it.  The other zeta functions ares calculated in below for the zero temperature and finite temperature.

\subsection{Zero Temperature : Kink-Antikink}
From the definition we have the relation

	$$\zeta_{V_1}(\nu) = [\Gamma(\nu)]^{-1} \int dx dt \int_0^{\infty}  ids (is)^{\nu -1}(-is) <x,t\mid e^{-isH_0} \left(-{1\over2}(1+x^2)x_0^2\right) \mid x,t> \hspace{2cm}$$
	$$ =[\Gamma(\nu)]^{-1} \int dx dt \int_0^{\infty} ds s^{\nu} \sum _n \int {p_0\over2\pi} <x,t\mid e^{-isH_0}\mid p_0, n><p_0, n \mid \left({1\over2}(1+x^2)\right) x_0^2\mid x,t>$$
	$$ = [\Gamma(\nu)]^{-1}  \sum _n \left[\int {p_0\over2\pi} \int_0^{\infty}  ds s^{\nu}  e^{-s (n+p_0^2)} \right]\left[\int dx \left({1\over2}(1+x^2)x_0^2\right) ~ D_n(x)^2\right]$$
	$$ = \nu \sum _n \left[\int {p_0\over2\pi} {1\over {(p_0^2 +n)^{\nu+1}}}\right] \left[(n+1)x_0^2\right]  \hspace{5cm}$$

	$$ = \nu x_0^2 ~ {\Gamma({1\over2})\Gamma(\nu +{1\over2})\over 2\pi \Gamma(\nu+1)} \sum _n {1+n\over {n^{\nu+{1\over2}}}} = \nu {x_0^2\over2}  ~\left[\zeta ({-1\over2})+\zeta ({1\over2}) \right]. \hspace{1.5cm}\eqno(3.1a)$$
\\
To obtain the above result  we have inserted the complete set $\mid p_0, n><p_0,n\mid  $ before the operator $H_0$ and replace $s$ by $-is$.  We have also transfered to the Euclidean space by $p_0 \rightarrow i p_0$.   Note that repeatedly using the recursion of the Hermite function $H_{n+1}(x) = 2x H_{n}(x)-2nH_{n-1}(x)$ we can expressed  $x^2 H_{n}$ as a linear combination of $H_{m}$ and then from the orthogonality of the Hermite function we have  a relation 

$$\int dx x^2 D_n(x) D_n(x) = \sqrt{(n+1)(n+2)} \delta_{n+2,m}+ (2n+1) \delta_{n,m}+\sqrt{(m+1)(m+2)} \delta_{n-2,m}.$$
\\
This formula has been used to obtain the result (3.1a).

   Now, substituting (3.1a) into (2.7) we can find that the bounded configuration of kink-antikink has the energy
 
 $$E_{K-\bar K} (x_0)  = -{1\over2} ~x_0^2~\left[\zeta ({-1\over2})+\zeta ({1\over2}) \right]  \approx  0.83 x_0^2. \eqno(3.2a)$$
\\
This result shows an attractive linear force between a kink and an antikink with short distance.  The contribution from the next order of $x_0^4$ could be evaluated in the same way [11].    While the first order have revealed the desired physical property we shall neglect it.

\subsection{Zero Temperature : Kink-Kink}
In the same way, for the  kink-kink system we have the relation

	$$\zeta_{V_1}(\nu) = [\Gamma(\nu)]^{-1} \int dx dt \int_0^{\infty}  ids (is)^{\nu -1}(-is) <x,t\mid e^{-isH_0} \left(-{1\over2}(1-x^2)x_0^2\right) \mid x,t> $$

	$$ =  \nu x_0^2 ~ {\Gamma({1\over2})\Gamma(\nu +{1\over2})\over 2\pi \Gamma(\nu+1)} \sum _n {- n\over {n^{\nu+{1\over2}}}} = \nu {x_0^2\over2}  ~\left[- ~ \zeta ({-1\over2}) \right]. \hspace{1.5cm}\eqno(3.1b)$$
\\
Thus the substituting the above equation (3.1b) into (2.7) we can find that the bounded configuration of kink-kink has the energy
 
 $$E_{K-K} (x_0)  = -{1\over2} ~x_0^2~\left[-~\zeta ({-1\over2}) \right]  \approx - 0.104 x_0^2. \eqno(3.2b)$$
\\
This result shows a repulsive force between a kink and a kink as desired.

\subsection{Finite Temperature}
To evaluate the zeta function for the kink-antikink at finite temperature we shall replace 

   $$p_0 \rightarrow ~i~{2\pi \over \beta} p_0,  \eqno(3.3)$$
    $$\int dp_0 \rightarrow ~i~{2\pi \over \beta} \sum_{p_0},  \eqno(3.4)$$
in which $p_0$ is an integral and temperature $T=\beta^{-1}$. Under these replacements the free energy $F$ is defined by [9]
    $$ \beta F = - {1\over 2} \zeta'(0).  \eqno(3.5) $$

Now following the prescription in the zero-temperature case we have the relation

   $$\zeta_{V_1}(\nu) =  [\Gamma(\nu)]^{-1} \int dx dt \int_0^{\infty} ds~ (s)^{\nu -1}(-s) <x,t\mid e^{-sH_0} \left(- {1\over2}(1 + x^2) x_0^4\right)\mid x,t> $$

   $$ = {\nu \over \beta} ~ x_0^2  \sum _{n=1} ~ \sum _{p_0} ~{1+n\over {\left[({2\pi p_0\over \beta})^2 +n \right]^{\nu+1}}}. \hspace{5cm} \eqno{(3.6)}$$
\\
To proceed let is first rewrite  the above equation as 

   $$  {\nu \over \beta} ~ x_0^2 ~ ({2\pi\over \beta})^{-2\nu-2 }
 \left[~2~ \sum _{n=1, p=1} {1+n\over {\left[p_0^2 + \left({\beta\over2\pi}\right)^2~n \right]^{\nu+1}}} + ... \right] $$

   $$ = {\nu \over \beta} ~ x_0^2 ~ ({2\pi\over \beta})^{-2\nu }
 \left[~2~ \sum _{ p=1}~\int dx ~ {1+ x~({2\pi\over \beta})^{2} \over {\left[p_0^2 + x \right]^{\nu+1}}} + ... \right] \hspace{-1cm}$$

  $$ = {\nu \over \beta} ~ x_0^2 ~ ({2\pi\over \beta})^{-2\nu }
 \left[~4~ \sum _{ p=1}~\int dy ~ {1+ y^2~({2\pi\over \beta})^{2} \over {\left[p_0^2 + y^2 \right]^{\nu+1}}} + ... \right] \hspace{-1cm}$$

  $$ = {2 \over \beta} ~ x_0^2 ~ \left[({2\pi\over \beta})^{-2\nu } ~\zeta(2\nu) 
- ({2\pi\over \beta})^{-2\nu +2} \zeta(2\nu-2) +... \right] \hspace{-1.5cm}$$  

$$ = {2 \over \beta} ~ x_0^2 ~ \left[~ln (kT)  + 2 \zeta(3)~ (kT)^2 +.. \right] \hspace{1.5cm}\eqno(3.7)$$
\\
in which the $...$ represent terms which are smaller compared to the leading term at high temperature, or those are $O(\nu^2)$ which do not contribute to the effective action.  To obtain the above result we have used a simple relation 

$$\int dy~{y^\beta\over \left(y^2+M^2 \right)^\alpha } = {\Gamma({1+\beta\over 2})\Gamma(\alpha -{1+\beta \over2})\over (M^2)^{\alpha -{1+\beta \over2}}\Gamma(\alpha) }\eqno{(3.8)}$$
\\
The reflation formula [14]
$$\pi^{-{z\over2}} ~ \Gamma ({z\over2}) ~ \zeta (z) = \pi^{-{1-z\over 2}}
~\Gamma ({1-z\over 2}) ~ \zeta (1-z) ,        \eqno{(3.9)}$$
has been used to regularize the divergence in the summation over $p_0$. (i.e., $\nu~\zeta(1+\nu) = 1 + \gamma \nu + ...$, in which $\gamma$ is the Eular constant) 

Substituting (3.7) into (3.5) we see that the free energy becomes local maximum at $x_0$ at high temperature, in contrast to the zero-temperature case in which the potential of kink-antikink system is minimum at $x_0=0$.   This indicates that the kink-antikink system may become stable at high temperature.

   Our next work is to find the transition temperature at which the zero-temperature phase (which is stable $x_0=0$)   is transfered to the high-temperature phase (which is stable at $x_0 \ne 0$).    However, as the stable distance at high temperature is very large, it is beyond the small $x_0$ approximation used in this paper.    Therefore let us  turn to investigate the low-temperature case.    

   At low temperature ${2\pi\over \beta} \ll 0$, we can use the Eular-Maclauriu formula 
$$\sum _k~ f(k) = \int dx f(x) - {1\over 12}~ f'(0) -  {1\over 720}~ f'''(0),   ~~~~~if~ f^{(n)}(\infty) = 0,\eqno{(3.10)}$$  
to take the summation over $p_0$ in (3.6).   The integration term becomes just that at zero term which have been presented at (3.1), and the $f'(0)$ terms is east to calculate. From (3.6)  and (3.1) the final result is

 $$\zeta_{V_1}(\nu)= {\nu \over \beta} ~ x_0^2  \sum _{n=1} ~ \sum _{p_0} ~{1+n\over {\left[({2\pi p_0\over \beta})^2 +n \right]^{\nu+1}}}. \hspace{6.5cm} $$

$$ \hspace{2.2cm}=\nu \sum _n \left[\int {p_0\over2\pi} {1\over {(p_0^2 +n)^{\nu+1}}}\right] \left[(n+1)x_0^2\right]  +  {\nu \over 6 \beta} ~ x_0^2~  ({2\pi \over \beta})^2 ~\left[\zeta(\nu +2) + \zeta(\nu +1)\right]$$

	$$= \nu~x_0^2 \left(~{1\over2}  ~\left[\zeta ({-1\over2})+\zeta ({1\over2}) \right] +~ {(2\pi)^2 ~(kT)^3}~\left[{1+\gamma\over6}+{1\over6} ~ \zeta (2)\right] \right)$$

  $$ \approx \nu~x_0^2~ \left(-1.33 + 0.617 (2\pi)^2 ~(kT)^3\right). \hspace{4.8cm} \eqno{(3.11)}$$
The transition temperature thus is at 
$$kT_c \approx \left({1\over 2 \pi^2}\right)^{1/3}. \eqno{(3.12)}$$
These complete our investigations.  We thus conclude that at the above critical temperature the kink-antikink system becomes stable. 

\section{Conclusion}

   In this paper we investigate the interactions within a parallel kink-kink and within a parallel kink-antikink.   We regard the pair of brane-antibrane as a configuration of kink or anti-kink which is a non-BPS brane and then use the tachyon effective field theory of non-BPS brane  [3,4,8] to investigate the interaction in this brane-antibrane systems.   We expand the Lagrangian of tachyon effective field theory to the  quadratic order in the off-diagonal fluctuation and use the zeta-function regularization and Schwinger perturbative formula  [9-11] to find the  to evaluate the interaction within a kink-kink or a kink-antikink.   The results show that while the kink and kink has repulsive force the kink and anti-kink has attractive force and may be annihilated by each others.   Especially, we also evaluate the free energy at finite temperature and determine the critical temperature above which the stable state of kink-antikink system may be found.

  Finally let us make two remarks: 

  (1)  Although the kinks or antikinks on type II brane-antibrane pairs correspond to unstable type II D-branes, which are uncharged, we can distinguish between the kink and the antikink (the notation used in this paper) in the following way.   Since the brane has not yet been annihilated by antibrane in a $D-\bar D$ pairs, there shall be a finite distance (which is assumed to be larger then the small distance $x_0$ used in eq.(2.11)) between the brane and antibrane.   Also, as brane has positive charge while antibrane  has negative charge  the brane-antibrane pairs can be, for example, regarded as an electric dipole system.     Now, for the brane-antibrane pairs along the x-axial we may call ($D-\bar D$) pairs a kink while ($\bar D-D$) pairs an antikink.   If a kink, i.e. a ($D_a-\bar D_a$) pairs is nearly overlapped on the another kink, i.e.  a ($D_b-\bar D_b$) pairs, then the repulsive force between $D_a$ and $D_b$, and repulsive force between $\bar D_a$ and $\bar D_b$ will lead to a repulsive force between kink and kink.   On the other hand, if a kink, i.e. a ($D_a-\bar D_a$) pairs,  is nearly overlapped on the another antikink, i.e. a ($\bar D_b-D_b$) pairs, then the attractive force between $D_a$ and $\bar D_b$, and attractive force between $\bar D_a$ and $D_b$ will lead to an attractive force between kink and antikink.

  (2)  Our treatment in this paper has used the Minahan-Zwiebach model [3,4] to analyze the kink-antikink system.   This model is a simplified field theory model of tachyon dynamics and is a powerful tool in describing various aspects of these dynamics in a simpler context.   It is known to share common properties with string theory.   However, the small distance expansions in eq.(2.11a) and eq.(2.11b)  is just a result of derivative truncation of BSFT [7].   It will be interesting to investigate the problem by using of an effective action (from string field theory) to describe a short distance computation and to see the validity of our procedure of integrating out the off diagonal terms of the tachyon, which seemingly still remain light in the short distance regime.  The problem with large $x_0$ is surely physical interesting and is also remained to be investigated.

\newpage

\begin{enumerate}
 \item A. Sen,  ``Tachyon Condensation on the Brane Antibrane System'' , 
  JHEP 9808  (1998)  012 ,   hep-th/9805170 ;    ``Descent Relations Among Bosonic D-branes'',  Int.\ J.\ Mod.\ Phys.  A14 (1999) 4061,  hep-th/9902105 ;   ``Non-BPS States and Branes in String Theory'',  hep-th/9904207; ``Universality of the Tachyon Potential'',  JHEP  9912 (1999) 027, hep-th/9911116 .
\item S.\ Moriyama and S.\ Nakamura,   ``Descent Relation of Tachyon Condensation from Boundary String  Field Theory'' ; Phys. Lett.  B506   (2001) 161,   hep-th/0009246 ; K. Hashimoto and S. Nagaoka,  ``Realization of Brane Descent Relations in Effective Theories'' , Phys. Rev  D66  (2002) 02060011,   hep-th/0202079 . 
\item  B.\ Zwiebach,   ``A Solvable Toy Model for Tachyon Condensation in String Field  Theory'' ,   JHEP  0009  2000  028 ,   hep-th/0008227 ; J.\ A.\ Minahan and B.\ Zwiebach,   ``Field Theory Models for Tachyon and Gauge Field String  Dynamics'' ,  JHEP  0009  2000  029 ,   hep-th/0008231 .  J.\ A.\ Minahan and B.\ Zwiebach,   ``Effective Tachyon Dynamics in Superstring Theory'' ,    JHEP  0103  2001  038 ,    hep-th/0009246 .  J.\ A.\ Minahan and B.\ Zwiebach,    ``Gauge Fields and Fermions in Tachyon Effective Field  Theories'' ,  JHEP  0102  2001  034 ,   hep-th/0011226 .
\item  D. Kutasov, M.\ Marino and G.\ Moore,  ``Remarks on Tachyon Condensation in Superstring Field  Theory'', hep-th/0010108; \ Minahan,   ``Stretched Strings in Tachyon Condensation Models'' ,  JHEP  0205  2002  024 , hep-th/0203108 .
\item  P.\ Kraus and F.\ Larsen,    ``Boundary String Field Theory of the  D$\bar  D $ System'' ,    PR  D63  2001  106004 ,      hep-th/0012198 ;  T.\ Takayanagi, S.\ Terashima and T.\ Uesugi,    ``Brane-Antibrane Action from Boundary String Field Theory'',  JHEP  0103  2001  019,  hep-th/0012210. 
\item K. Hashimoto and N. Sakai, ``Brane - Antibrane as a Defect of Tachyon Condensation '' ,  hep-th/0209232 . 
\item  E. Witten,  ``On Background Independent Open String Field Theory'' ,   Phys. Rev.   D46  (1992) 5467,  hep-th/9208027;   ``Some Computations in Background Independent Off-Shell String  Theory'' , Phys. Rev. D47 (1993)  3405, hep-th/9210065;  A. A. Gerasimov and S. L. Shatashvili,    ``On Exact Tachyon Potential in Open String Field Theory'' ,    JHEP  0010  (2000)  034 ,   hep-th/0009103 ; D. Kutasov, M.\ Marino and G.\ Moore,  ``Some Exact Results on Tachyon Condensation in String Field  Theory'',  JHEP  0010  (2000)  045, hep-th/0009148; V.\ Niarchos and N.\ Prezas, ``Boundary Superstring Field Theory'',  Nucl. Phys.  B619 (2001)  51,  hep-th/0103102; O.\  Andreev,    ``Some Computations of Partition Functions and Tachyon  Potentials in Background Independent Off-Shell String Theory'',  Nucl. Phys.  B598  (2001)  151,  hep-th/0010218 . 
\item  A.~A.~Gerasimov and S.~L.~Shatashvili, ``On non-abelian structures in field theory of open strings,'' JHEP {\bf 0106}, 066 (2001) [arXiv:hep-th/0105245]; E.~T.~Akhmedov, ``Non-Abelian structures in BSFT and RR couplings'',  hep-th/0110002. V.~Pestun, ``On non-Abelian low energy effective action for D-branes'', JHEP {\bf 0111}, 017 (2001), hep-th/0110092.
\item  S. Hawking, Commum. Math. Phys. 55, 133 (1976)
\item   J. Schwinger, Phys. Rev. 82, 664 (1951);  L. Parker, " Aspects of quantum filed theory in curved spacetime: effective action and energy-momentum tensor" in "Recent Developments in Gravitation", ed. S. Deser and M. Levey (New York: Plenum, 1977).
\item  W. H. Huang, One-loop Effective Action on Rotational Spacetimes: $\zeta$ - Function Regularization and Swinger Perturbative Expansion, Ann. Phys. 254, 69 (1997), hep-th/0211079; W. H. Huang,  "Semiclassical gravitation and quantization for the Bianchi type I universe with large anisotropy ", Phys. Rev. D 58 (1998) 084007, hep-th/0209092;  W. H. Huang, Conformal Transformation of Renormalized Effective in Curved Spacetimes, Phys. Rev. D 51, 579 (1995).  
\item U. H. Danielsson, A.Guijosa, and M. Kruczenski, " Brane-Antibrane Systems at Finite Temperature and the Entropy of Black Branes", JHEP 0109 (2001) 011, hep-th/00106201; M. Majumdar and A. Davis, "Cosmological Creation of D-branes and anti-D-branes" , JHEP 0203 (2002) 056 , hep-th/0202148.
\item    I. S. Gradshteyn and I. M. Ryzhik ,"Table of Intergals, Series and Products." Academic Press. New York 1980.
\item J. Ambjorn and S. Wolfram, Ann. Phys. {\bf 147} (1983) 1; E. Elizalde, S. d. Osdinsov, A. Romoeo, A. A. Bytseko, and S. Zerbini, "Zeta Regularization Techniques with Applications", World Scientific, 1994.

\end{enumerate}
\end{document}